\documentclass[aps,prd,preprint,superscriptaddress]{revtex4-1}

\usepackage{lineno}
\usepackage{color}
\usepackage{graphicx}
\usepackage{hyperref}

\newcommand{\jpsi}{\mathrm{J/}\psi}
\newcommand{\psip}{\psi(2\mathrm{S})}

\newcommand{\Wgp}{W_{\gamma\mathrm{p}}}

\newcommand{\Rz}{\rho^{0}}
\newcommand{\Uos}{\Upsilon(1S)}
\newcommand{\Uts}{\Upsilon(2S)}

\modulolinenumbers[5]
\begin{document}

%\preprint{}


\title{
Dissociative production of vector mesons at  electron-ion colliders
}


\author{D. Bendova}
\affiliation{Faculty of Nuclear Sciences and Physical Engineering,
Czech Technical University in Prague, Czech Republic}
\author{J. Cepila}
\affiliation{Faculty of Nuclear Sciences and Physical Engineering,
Czech Technical University in Prague, Czech Republic}
\author{J. G. Contreras}
\affiliation{Faculty of Nuclear Sciences and Physical Engineering,
Czech Technical University in Prague, Czech Republic}


\date{\today}

\begin{abstract}
We present predictions for the exclusive and  dissociative production of  vector mesons off  protons in an electron-ion collider. The computation is based on the energy-dependent hot spot model that was shown to  successfully describe the available photoproduction data. We find that the model also describes correctly all available electroproduction data. In addition, we find that the cross section for dissociative production as a function of the center-of-mass energy of the photon-proton system has a maximum, whose position depends on the virtuality of the photon and the mass of the vector meson. We use these maxima to define a geometrical saturation scale and find that it grows linearly with energy as a function of the scale of the process. This phenomenon can be studied at the proposed electron-ion colliders, JLEIC, eRHIC and LHeC.

\end{abstract}

% 12.38.-t	Quantum chromodynamics 
% 13.60.-r	Photon and charged-lepton interactions with hadrons 
% 13.60.Le	Meson production 

\maketitle

\section{Introduction
\label{sec:intro}}

Within perturbative Quantum Chromodynamics (pQCD), the structure of hadrons in terms of its constituent partons evolves with  energy, or equivalently with Bjorken-$x$. Very precise measurements of the $F_2(x,Q^2)$ structure function of the proton performed at HERA with photons of virtuality $Q^2$ indicate that the gluon density grows steeply for decreasing $x$~\cite{Abramowicz:2015mha}.
According to pQCD this behavior changes at some point where non-linear effects start to be important and the proton structure enters a regime known as saturation; see for example~\cite{Gelis:2010nm} and references therein.

\begin{figure*}[t!]
\includegraphics[width=0.48\textwidth]{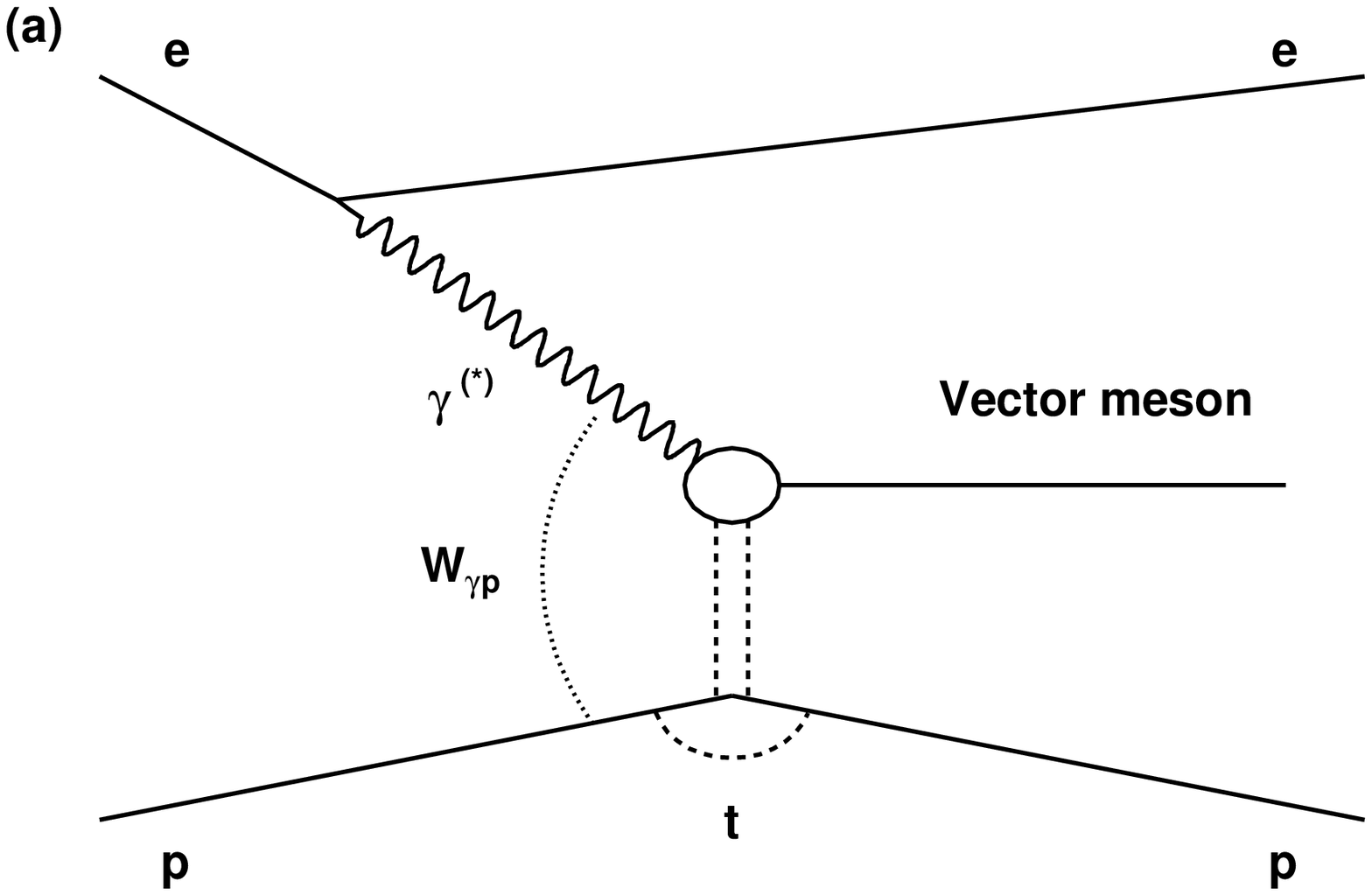}
\includegraphics[width=0.48\textwidth]{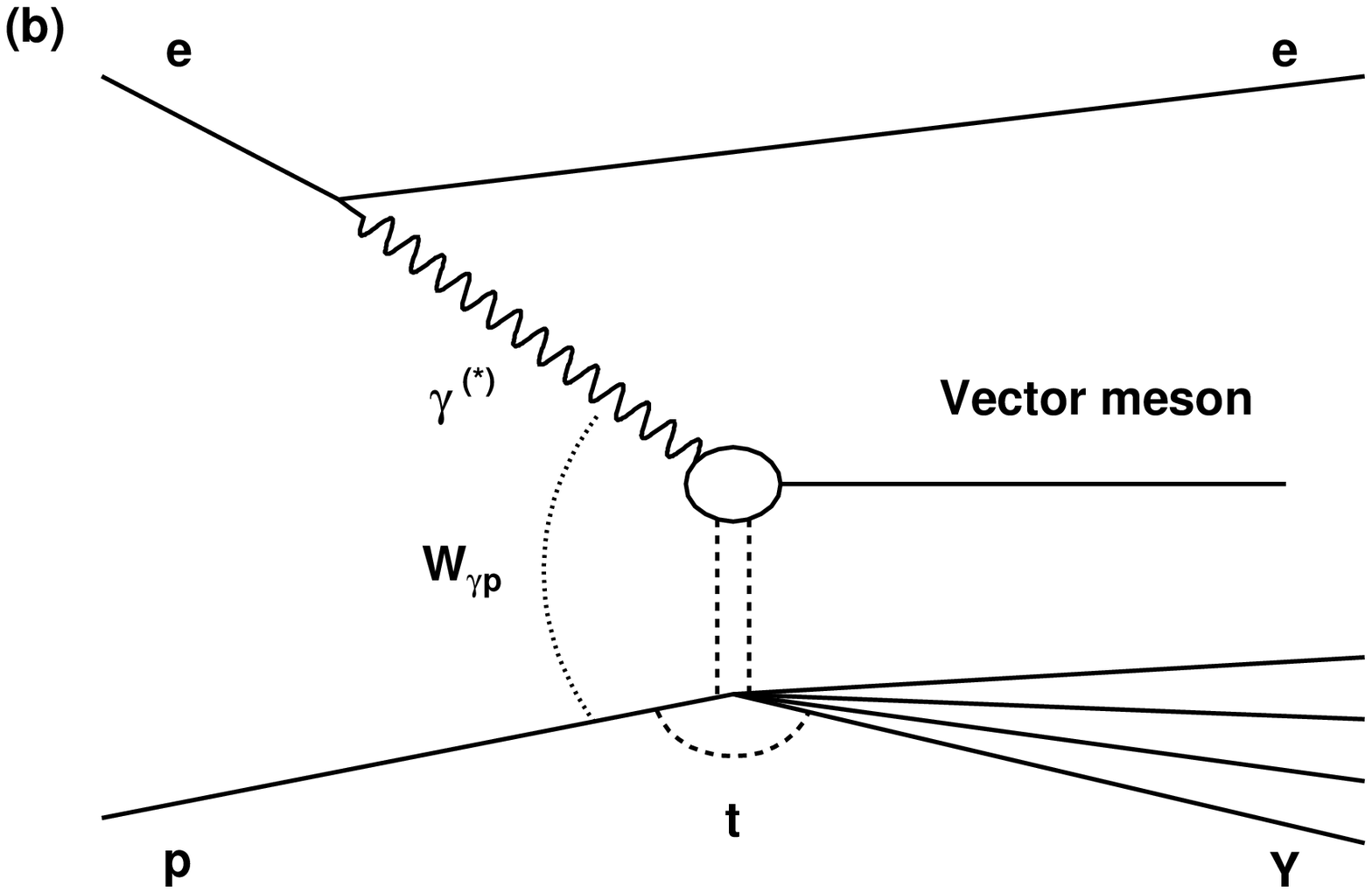}
\caption{\label{fig:diag}  Diagrams for exclusive ($a$) and dissociative ($b$)
production of vector mesons in an electron-ion collider. See text for details.}
\end{figure*}

Exclusive vector meson production in electron-hadron colliders, depicted in Fig.~\ref{fig:diag} (a), has been advocated as a tool to study the saturation phenomenon in the  facilities that are under design now, like the EIC or the LHeC~\cite{Accardi:2012qut,AbelleiraFernandez:2012cc}. In this process, the incoming electron emits a photon which interacts with the proton to produce a vector meson. The photon can be quasi-real ($\gamma$) or have a large virtuality ($\gamma^*$); these cases are known as photo- or electroproduction, respectively. Here, $\Wgp$ is the center-of-mass energy of the photon-proton system and $-t$ is the square of the momentum transferred in the proton vertex. 
This process has been extensively investigated at HERA and at the LHC. (For recent reviews see~\cite{Newman:2013ada} and~\cite{Contreras:2015dqa}, respectively.) These measurements have been  successfully described by a variety of models including saturation effects; e.g.~\cite{Kowalski:2006hc,Armesto:2014sma,Goncalves:2014wna}. 
A recent study addresses in detail the corresponding measurements at future electron-ion colliders~\cite{Lomnitz:2018juf}.
 
 A related process, shown schematically in Fig.~\ref{fig:diag} (b), that has recently attracted renewed attention, is the production of a vector meson accompanied by the dissociation of the scattered proton. In a Good-Walker approach~\cite{Good:1960ba, Miettinen:1978jb} this process can be related to fluctuations of the partonic structure of the proton~\cite{Mantysaari:2016ykx,Mantysaari:2016jaz}. Specifically, it is related to the variance over the different configurations of the partonic structure, and the main contribution to the variance is given by fluctuations in the geometrical configurations in the impact-parameter plane. Using a model with three so-called hot spots --- regions of high gluonic density ---, the authors of~\cite{Mantysaari:2016ykx} showed that the measurement of the cross section for the dissociative photoproduction of $\jpsi$  as a function of $|t|$, at a fixed $\Wgp$, could be successfully described.
 
 These ideas were extended in~\cite{Cepila:2016uku} by the inclusion of an energy dependence on the number of hot spots, which grows with decreasing $x$, mimicking the expectations of pQCD. This model successfully describes all available data on the energy dependence of both exclusive and dissociative photoproduction of $\jpsi$ off protons. Furthermore, it predicts that the dissociative cross section grows with energy up to a maximum value and then decreases steeply. These investigations were continued in~\cite{Cepila:2017nef} and~\cite{Cepila:2018zky} to describe the production off nuclear targets and of different vector mesons, respectively. In~\cite{Cepila:2018zky}, it was observed that the position of the maximum of the dissociative cross section depends on the mass of the vector meson in photoproduction processes.
 
In this article, we apply our model to the case of the dissociative electroproduction of vector mesons. 
We find that this cross section has a maximum, whose position depends on the virtuality of the photon and the mass of the vector meson. We use these maxima to define a geometrical saturation scale and find that it grows linearly with energy as a function of the scale of the process, as reported  in Fig~\ref{fig:mass-Q_dep}. The rest of this contribution is organized as follows. A brief description of the formalism is presented in Sec.~\ref{sec:form}. The model predictions are presented and compared to the available data in Sec.~\ref{sec:res}. Section~\ref{sec:gss} introduces the geometrical saturation scale. We close with a brief summary and outlook in Sec.~\ref{sec:sum}.
 
\section{Description of the formalism
\label{sec:form}}

\subsection{The photon-proton scattering amplitude}
The diffractive production of a vector meson when a virtual photon interacts with a proton can be well described within the color dipole picture \cite{Mueller:1993rr,Mueller:1994jq}. In this case, the scattering amplitude takes the following form (for a detailed derivation see e.g.~\cite{Kowalski:2006hc}),

\begin{widetext}
\begin{equation}
\mathcal{A}_{T,L}(x,Q^2,\vec{\Delta}) = i \int \mathrm{d}\vec{r} \int \limits_0^1 \frac{\mathrm{d}z}{4\pi} \int \mathrm{d}\vec{b} |\Psi_{\rm V}^* \Psi_{\gamma^*}|_{T,L} \exp \left[ -i\left( \vec{b} - (1-z)\vec{r} \right)\vec{\Delta} \right] \frac{\mathrm{d}\sigma_{q\bar{q}}}{\mathrm{d} \vec{b}},
\label{VM-amplitude}
\end{equation}
\end{widetext}
where the subscripts $T$ and $L$ denote the contribution from the transversally, respectively longitudinally, polarized virtual photon. $\Psi_{\rm V}$ is the wave function of the vector meson, $\Psi_{\gamma^*}$ is the wave function of a virtual photon, which fluctuates into a quark-antiquark dipole, $\vec{r}$ is the transverse size of the color dipole, $z$ is the fraction of the photon longitudinal momentum carried by the quark, $\vec{b}$ is the impact parameter and $\vec{\Delta}^2 \equiv-t$. The Bjorken-$x$ of the exchanged pomeron is, under the assumption of large $\Wgp$, given by
\begin{equation}
x = \frac{Q^2 + M^2}{\Wgp^2 + Q^2},
\label{x}
\end{equation}
with $M$ being the invariant mass of the given vector meson. Finally, $\mathrm{d}\sigma_{q\bar{q}}/\mathrm{d} \vec{b}$ is the cross section for the interaction of the color dipole and the target.

In this formalism, the exclusive cross section  to produce the vector meson V is given by
\begin{equation}
\frac{\mathrm{d}\sigma^{\gamma^*p \rightarrow {\rm V}p}}{\mathrm{d}|t|} \bigg| _{T,L} = \frac{\left(R_g ^{T,L}\right)^2}{16\pi} | \langle \mathcal{A}_{T,L} \rangle |^2,
\label{VM-cs-diff-excl}
\end{equation}
while the cross section where the proton dissociates into a system $Y$ is
\begin{equation}
\frac{\mathrm{d}\sigma^{\gamma^*p \rightarrow {\rm V}Y}}{\mathrm{d}|t|} \bigg| _{T,L} = \frac{\left(R_g ^{T,L}\right)^2}{16\pi} \left( \langle |\mathcal{A}_{T,L}|^2 \rangle - | \langle \mathcal{A}_{T,L}  \rangle|^2 \right).
\label{VM-cs-diff-disoc}
\end{equation}
In both cases, the total cross section is given by the sum of  the transverse and the longitudinal contributions.
The factor $R_g ^{T,L}$ is called the skewedness correction~\cite{Shuvaev:1999ce} and takes into account that there are two values of $x$ involved in the interaction but only one appears in Eq.~(\ref{VM-amplitude}).

There are two ingredients of Eq.~(\ref{VM-amplitude}) that need to be modeled: the wave function to create a vector meson out of the quark-antiquark dipole and the cross section for the interaction of the color dipole and the target. They are discussed in the following.

\subsection{Wave functions of vector meson }

The wave functions of vector mesons  are modeled assuming that the vector meson is predominantly a $q\bar{q}$ pair with the same polarization structure as the photon. The overlap of the photon-meson wave functions in Eq.~(\ref{VM-amplitude}) is given as 
\begin{widetext}
\begin{equation}
|\Psi_{\rm V} ^* \Psi_{\gamma^*}|_T = \hat{e}_f e \frac{N_C}{\pi z(1-z)} \left[ m_f ^2 K_0 (\epsilon r) \phi_T (r,z) - \left( z^2 + (1-z)^2 \right) \epsilon K_1 (\epsilon r) \partial_r \phi _T (r,z) \right],
\label{VM-psipsiT}
\end{equation}
and
\begin{equation}
|\Psi_{\rm V} ^* \Psi_{\gamma^*}|_L = \hat{e}_f e \frac{N_C}{\pi}2Qz(1-z) K_0(\epsilon r) \left[ M \phi _L (r,z) + \delta \frac{m_f ^2 - \nabla _r ^2}{Mz(1-z)} \phi _L (r,z) \right],
\label{VM-psipsiL}
\end{equation}
\end{widetext}
where  $r\equiv|\vec{r}|$, $N_C$ is the number of colors, the (effective) mass of the given flavor is $m_f$,  and an effective charge  is denoted by $\hat{e}_f$. The parameter $\delta$ is a switch to include or not the corresponding term; we set it equal to one, which corresponds to the boosted Gaussian model~\cite{Nemchik:1994fp, Nemchik:1996cw, Forshaw:2003ki}. $K_i$ are Bessel functions and
\begin{equation}
\epsilon = z(1-z)Q^2 + m_f ^2.
\end{equation}

The scalar part $\phi _{T,L}$ of the wave function is in general model-dependent.  In the boosted Gaussian model the scalar part is described by the Gaussian distribution
\begin{widetext}
\begin{equation}
\phi_{T,L}(r,z)=N_{T,L}z(1-z)\exp\left(-\frac{m_f^{2}R^{2}}{8z(1-z)}-\frac{2z(1-z)r^{2}}{R^{2}}+\frac{m_f^{2}R^{2}}{2}\right).
\end{equation}
The parameters of the model are fixed using a normalization condition and the measured electronic decay width (see, e.g.~\cite{Kowalski:2006hc}) . For the first excited state 2S, the scalar wave function has the form 
\begin{equation}
\phi^{\rm 2S}_{T,L}(r,z)=\Phi_{T,L}(r,z)\left( 1+\alpha_{\rm 2S}\left(2+\frac{m_f^{2}R^{2}}{4z(1-z)}-\frac{4z(1-z)r^{2}}{R^{2}}-m_f^{2}R^{2}\right)\right).
\end{equation}
\end{widetext}
The condition that the 1S and 2S states are orthogonal, fixes the extra parameter $\alpha_{\rm 2S}$. 

We have recomputed the values of the parameters for the wave functions of all vector mesons discussed in the following to match them to the measurements gathered in the PDG of 2016~\cite{Patrignani:2016xqp}. The parameter values are reported in Tab.~\ref{Tab:mesons}. 

\begin{table*}%[H] add [H] placement to break table across pages
\caption{\label{Tab:mesons} Parameters for vector meson  (V) wave functions: mass of the vector meson $M$, effective mass of the given flavor $m_f$, effective charge $\hat{e}_f$, scalar part parameters $N_T$, $N_L$, $R^2$ and $\alpha_{\rm 2S}$, fixed with the values reported in the 2016 PDG~\cite{Patrignani:2016xqp}. }
\begin{ruledtabular}
%\begin{tabular}{|c||c|c|c|c|c|c|c|}
\begin{tabular}{cccccccc}
V & $M \; [\mathrm{GeV}$] & $m_f \; [\mathrm{GeV}]$ & $\hat{e}_f \; [-]$ & $N_T \; [-]$ & $N_L \; [-]$ & $R^2 \; [\mathrm{GeV^{-2}}]$ & $\alpha_{\rm 2S} \; [-]$ \\
\hline
$\Rz$ & 0.775260 & 0.14 & $1/\sqrt{2}$ & 0.909 & 0.853 & 12.75 & -- \\
%\hline
$\phi$ & 1.019461 & 0.14 & 1/3 & 0.918 & 0.823 & 11.3 & -- \\
%\hline
$\jpsi$ & 3.09690 & 1.4 & 2/3 & 0.582 & 0.578 & 2.24 & -- \\
%\hline
$\psip$ & 3.686097 & 1.4 & 2/3 & 0.666 & 0.658 & 3.705 & -0.6225 \\
%\hline
$\Uos$ & 9.46030 & 4.2 & 1/3 & 0.478 & 0.478 & 0.585 & -- \\
%\hline
$\Uts$ & 10.02326 & 4.2 & 1/3 & 0.614 & 0.610 & 0.831 & -0.568 \\
\end{tabular}
\end{ruledtabular}
\end{table*}

\subsection{Dipole-target cross section}

 The cross section for the interaction between the color dipole with the proton target is related, via the optical theorem, to the imaginary part of the dipole-proton amplitude $N(x,\vec{r},\vec{b})$:

\begin{equation}
\frac{\mathrm{d} \sigma_{q\bar{q}}}{\mathrm{d}\vec{b}} = 2 N(x,\vec{r},\vec{b}) .
\label{VM-dipole-cs}
\end{equation}

In order to separate the effects of fluctuations of the proton structure in the transverse plane  from the energy dependence of the cross section we proposed in~\cite{Cepila:2016uku} to use the factorized form 
\begin{equation}
\frac{\mathrm{d} \sigma_{q\bar{q}}}{\mathrm{d}\vec{b}} = \sigma_0 N(x,r)T_p(\vec{b}),
\end{equation}
where $T_p(\vec{b})$ decribes the proton profile in the impact-parameter plane and $\sigma_0$ is a normalization parameter, which we fixed to $\sigma_0 = 4\pi B_p$. The interpretation of $B_p$ is discussed below.

The dipole amplitude $N(x,r)$ can be obtained from various parameterizations (for an overview see e.g. \cite{Kowalski:2006hc}) or as the solution of the Balitsky-Kovchegov evolution equation \cite{Balitsky:1995ub,Kovchegov:1999yj}. To keep the model as simple as possible, we have chosen the form of the dipole amplitude $N(x,r)$ given by the Golec-Biernat and Wusthoff model \cite{GolecBiernat:1998js,GolecBiernat:1999qd}, 

\begin{equation}
N(x,r) = \left[ 1 - \exp \left( -\frac{r^2 Q_s ^2 (x)}{4} \right) \right],
\label{VM-dipole-cs-GBW}
\end{equation}
where $Q_s(x)$ is the so-called  saturation scale, which in this model is given by
\begin{equation}
\label{eq:q2s}
Q_s ^2 (x) = Q_0 ^2 \left( \frac{x_0}{x} \right) ^{\lambda}.
\end{equation}

Since the proton is a quantum object, its structure changes from interaction to interaction. To incorporate this effect we use a model of the proton as constituted by hot spots ($hs$), which represent regions of high gluon density. The positions of these hot spots in the transverse plane  fluctuate event-by-event and are described by  the proton profile function $T_p (\vec{b})$, which is defined as 
\begin{equation}
T_p(\vec{b}) = \frac{1}{N_{hs}} \sum \limits_{i=1}^{N_{hs}} T_{hs} \left( \vec{b} - \vec{b}_i \right),
\label{VM-hs-T}
\end{equation}
where each hot spot is defined as
\begin{equation}
T_{hs} (\vec{b} - \vec{b}_i) = \frac{1}{2\pi B_{hs}} \exp \left( -\frac{\left( \vec{b} - \vec{b}_i\right)^2}{2B_{hs}} \right).
\label{VM-hs}
\end{equation}
Each vector $\vec{b_i}$ is obtained from a two-dimensional Gaussian distribution with width $B_p$ and centered at (0,0). Thus, the parameters $B_p$ and  $B_{hs}$ can be interpreted as half of the average of the squared radius of the proton and of the hot spot, respectively. In this sense $\sigma_0 = 4\pi B_p$ is a measure of the overall transverse area of the proton.

The key feature of our model is the evolution of the number of hot spots with energy.  $N_{hs}$ is
a random number drawn from a zero-truncated Poisson distribution, where the Poisson distribution has a mean value
\begin{equation}
\langle N_{ hs}(x) \rangle = p_0x^{p_1}(1+p_2\sqrt{x}),
\label{eq:Nhsx}
\end{equation}
where $p_0$, $p_1$ and $p_2$ are parameters.

The values of all parameters of our model were fixed in earlier publications~\cite{Cepila:2016uku,Cepila:2017nef,Cepila:2018zky} using $\jpsi$ data from photoproduction at HERA. The values  are listed here for completeness:  $B_p=4.7$ GeV$^{-2}$, $B_{hs}=0.8$ GeV$^{-2}$, $p_0 = 0.011$, $p_1 = -0.58$, $p_2=300$, $\lambda=0.21$, $x_0=2\times10^{-4}$ and $Q_0=1$ GeV. In order to describe the normalization of the photoproduction of $\rho$ and of $\phi$ we set $B_p=8$ GeV$^{-2}$ as done in ~\cite{Cepila:2018zky} and consistent with the observations of the H1 Collaboration~\cite{Aaron:2009xp}. For the case of electroproduction discussed below we set $B_p=4.7$ GeV$^{-2}$ for all vector mesons.

\section{Predictions and comparison with experimental data
\label{sec:res}}

Using the model described above we predict the energy dependence of the exclusive and dissociative production of vector mesons off a proton target for $\Rz$, $\phi$, $\jpsi$, $\psip$, $\Uos$ and $\Uts$ at different virtualities of the exchanged photon. We compare our predictions with data when available. For completeness we also show the predictions for photoproduction that were presented in~\cite{Cepila:2018zky}.

\begin{figure*}
\includegraphics[width=0.48\textwidth]{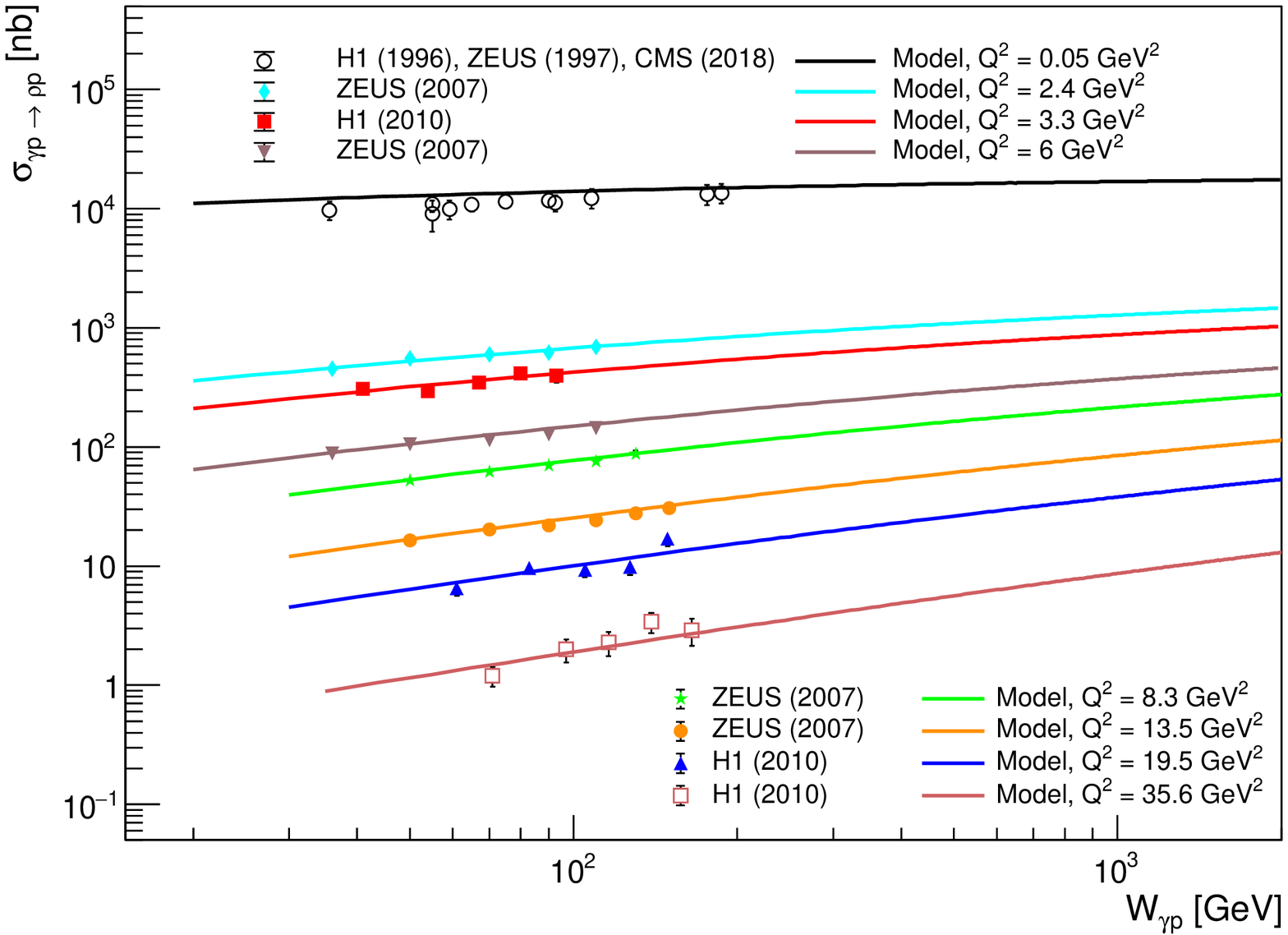}
\includegraphics[width=0.48\textwidth]{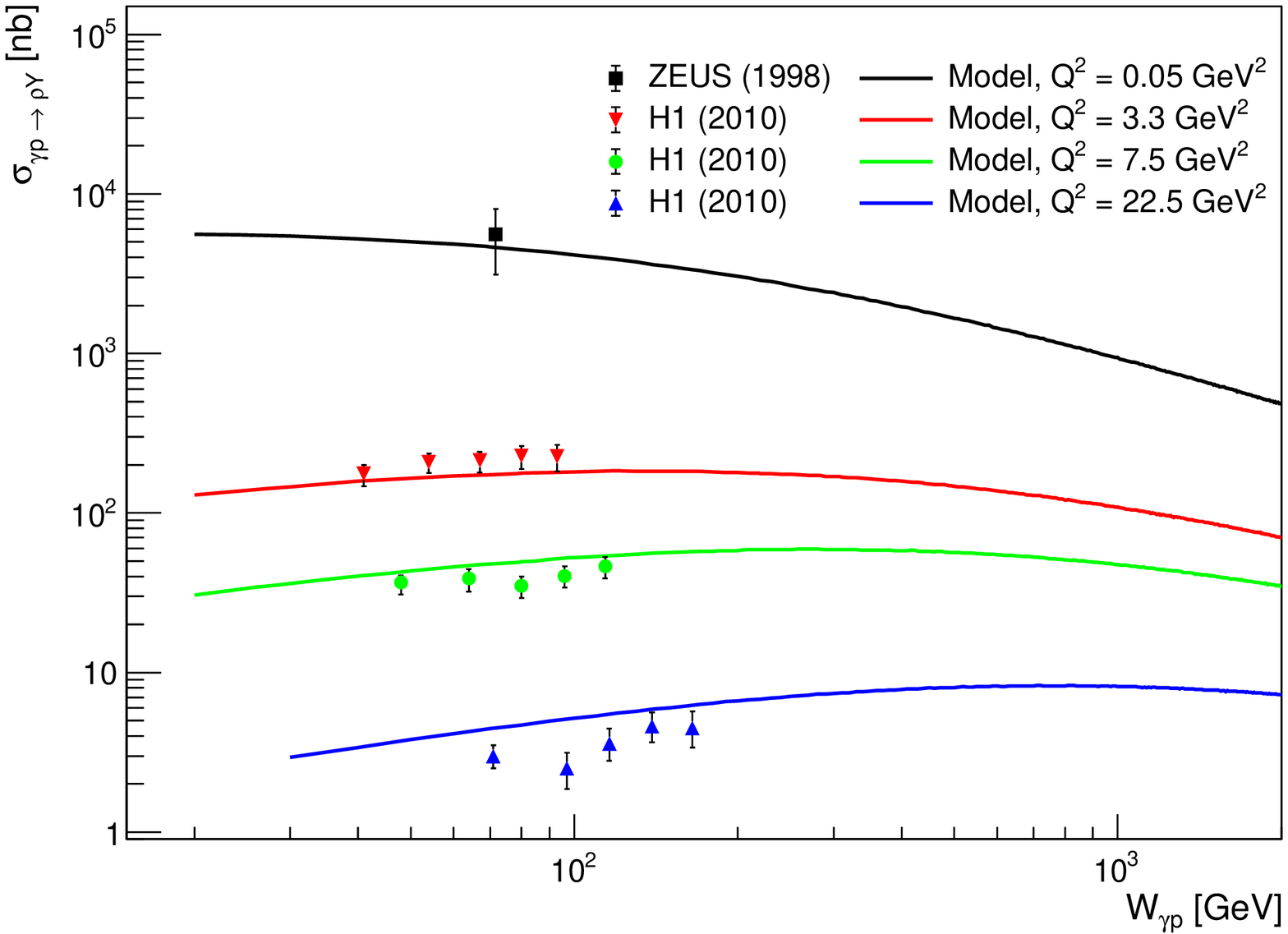}
\caption{\label{fig:rho} (Color online) Comparison of the model predictions (solid lines) with  HERA~\cite{Aaron:2009xp,Aid:1996bs,Adloff:1997jd,Breitweg:1997ed,Chekanov:2007zr} and CMS data~\cite{CMS-PAS-FSQ-16-007} for the  $\Wgp$ dependence of the exclusive (left) and dissociative (right) photo- and electroproduction  cross section of a $\Rz$ meson.}
\end{figure*}

\begin{figure*}
\includegraphics[width=0.48\textwidth]{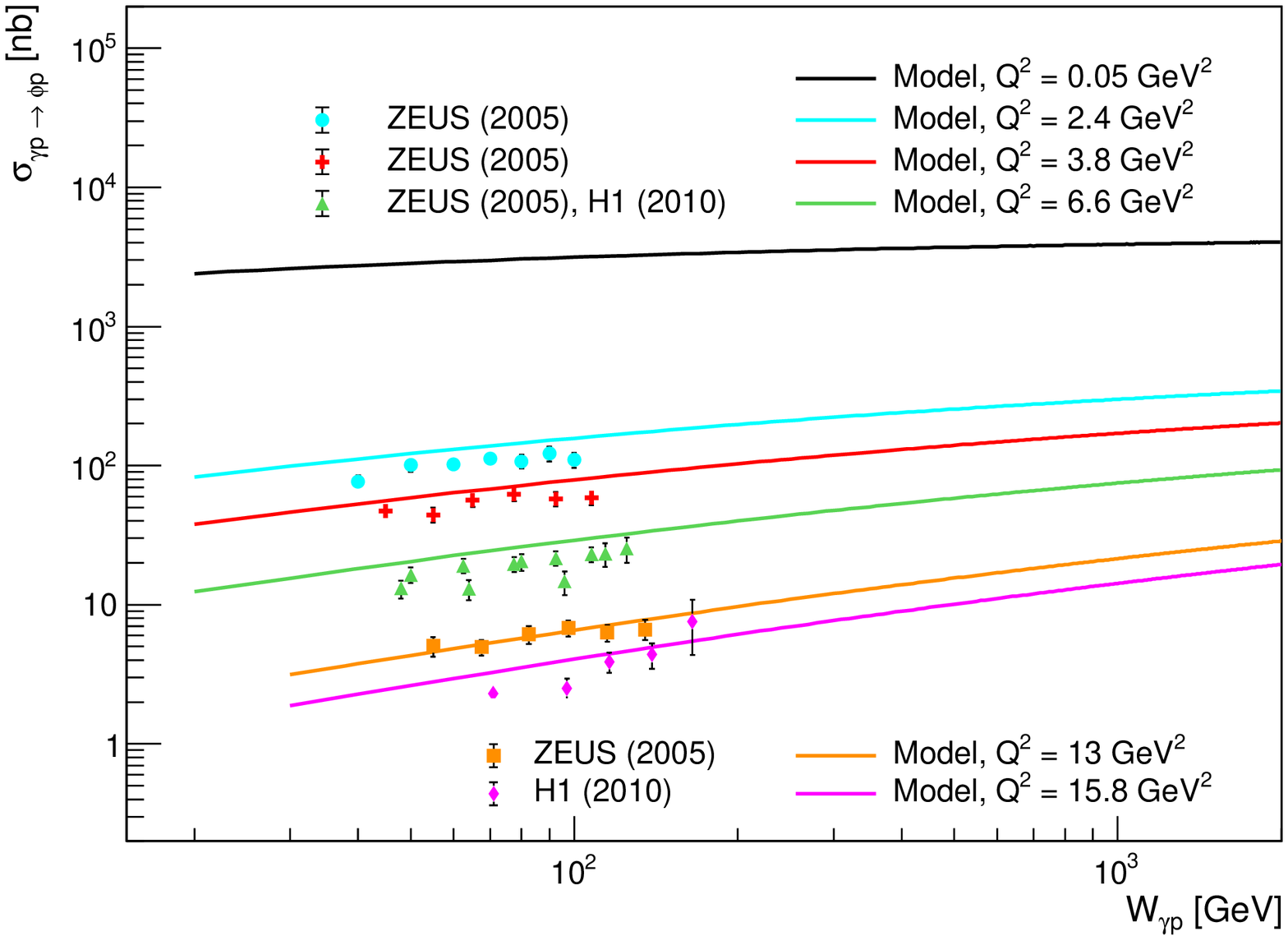}
\includegraphics[width=0.48\textwidth]{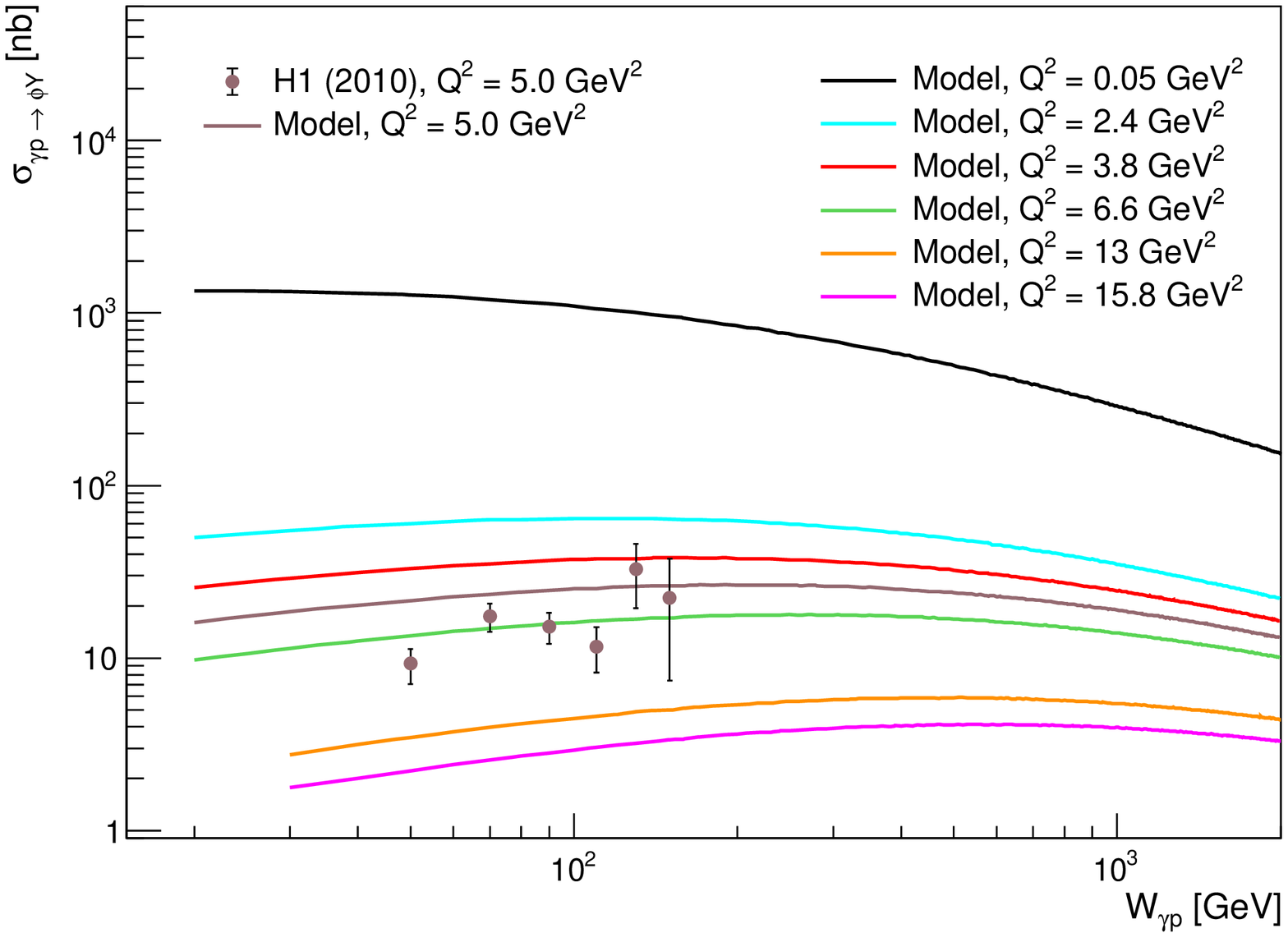}
\caption{\label{fig:phi} (Color online) Comparison of the model predictions (solid lines) with  HERA data from H1~\cite{Aaron:2009xp,Adloff:1997jd} and ZEUS~\cite{Chekanov:2005cqa} for the  $\Wgp$ dependence of the exclusive (left) and dissociative (right) photo- and electroproduction  cross section of a $\phi$ meson.}
\end{figure*}

The predictions for the $\Wgp$ dependence of the exclusive and dissociative cross section of the $\Rz$ vector meson are presented in Fig. \ref{fig:rho}. Predictions are compared with H1~\cite{Aid:1996bs,Adloff:1997jd,Aaron:2009xp} and ZEUS data~\cite{Breitweg:1997ed,Chekanov:2007zr} for several values of $Q^2$ and also to the preliminary CMS data \cite{CMS-PAS-FSQ-16-007} for photoproduction in p--Pb collisions at the center-of-mass energy $\sqrt{s} = 5.02$  TeV. The predictions for electroproduction, both exclusive and dissociative, give a very good description of the available data covering virtualities from 2.4 GeV$^2$ to 35.6 GeV$^2$. Recently, the H1 Collaboration released preliminary data (not shown in the figure) on the energy dependence of $\Rz$ dissociative photoproduction. The predictions of our model are consistent with these preliminary data, although a definitive comparison can only be done after the measurement is published in its final form. 

The predictions for the energy-dependence of the exclusive and dissociative  photo- and electroproduction cross sections of the $\phi$ vector meson are compared with  H1~\cite{Adloff:1997jd,Aaron:2009xp} and ZEUS data~\cite{Chekanov:2005cqa} in Fig. \ref{fig:phi}. The description of the electroproduction data is satisfactory, however it is not as good as for the case of the $\Rz$ meson. 

\begin{figure*}
\includegraphics[width=0.48\textwidth]{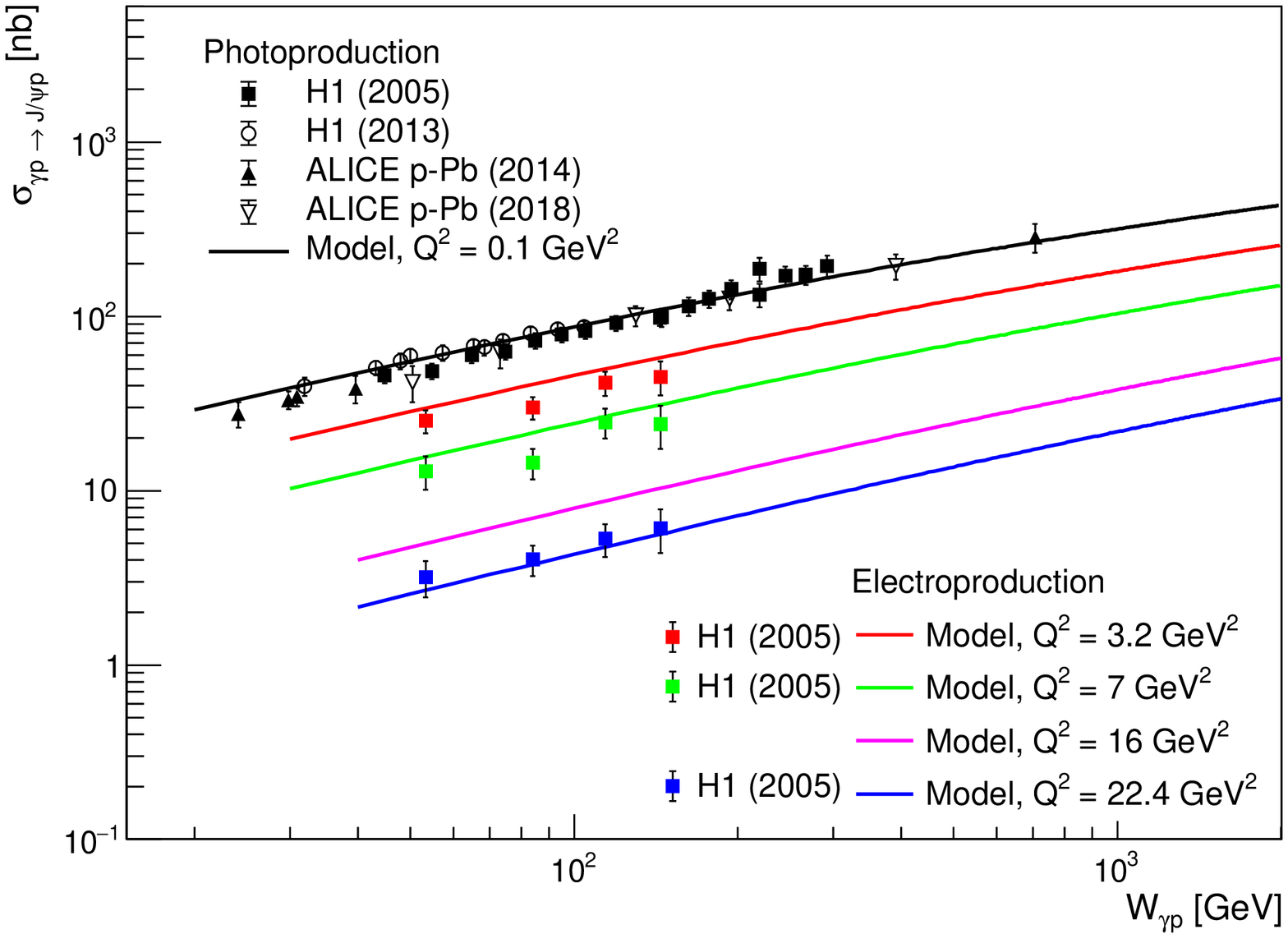}
\includegraphics[width=0.48\textwidth]{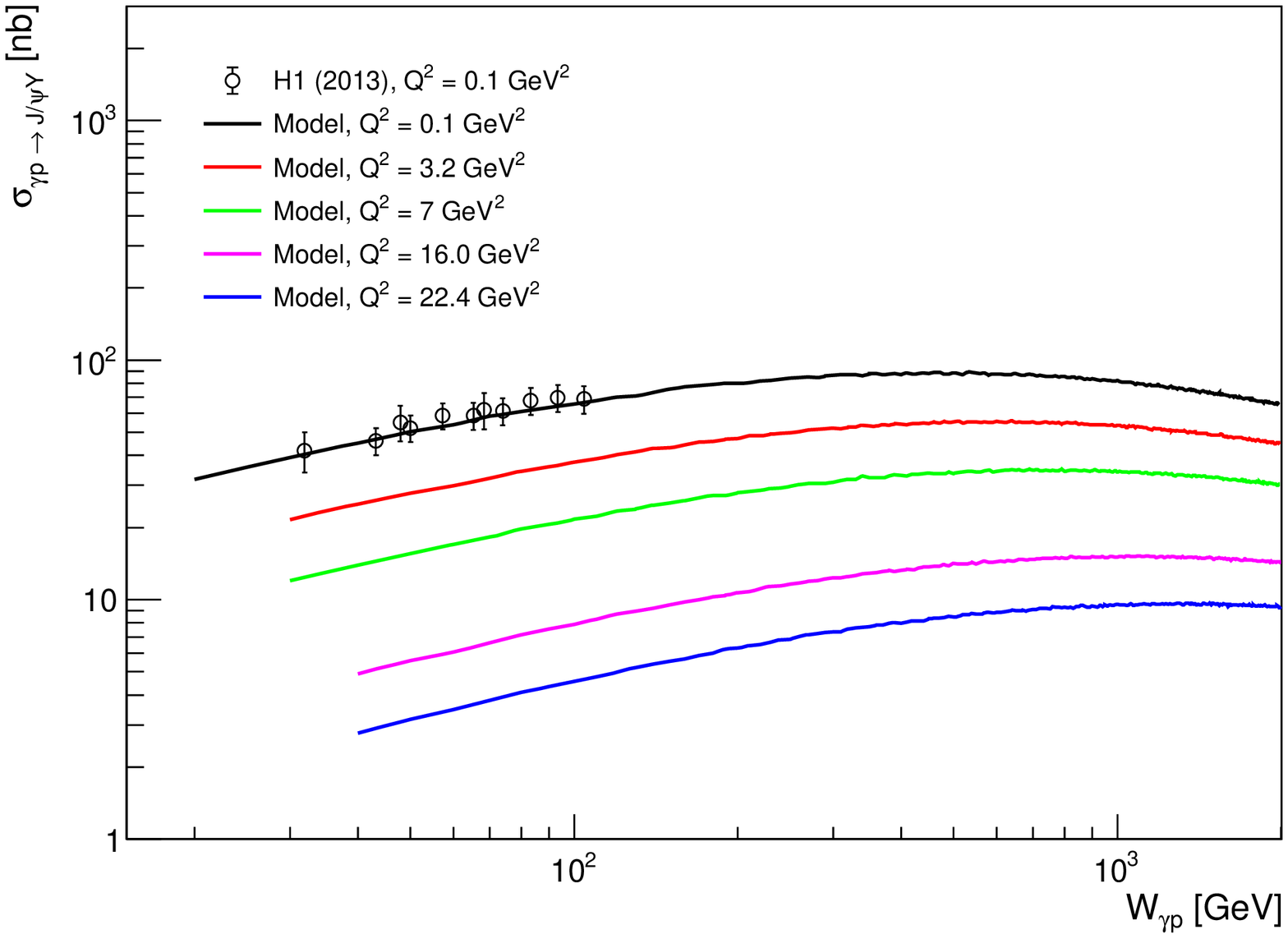}
\caption{\label{fig:J-psi} (Color online) Comparison of the model predictions (solid lines) with H1~\cite{Aktas:2005xu,Alexa:2013xxa} and ALICE data~\cite{TheALICE:2014dwa,Acharya:2018jua} for the  $\Wgp$ dependence of the exclusive (left) and dissociative (right) photo- and electroproduction  cross section of a $\jpsi$ meson.}
\end{figure*}

\begin{figure*}
\includegraphics[width=0.48\textwidth]{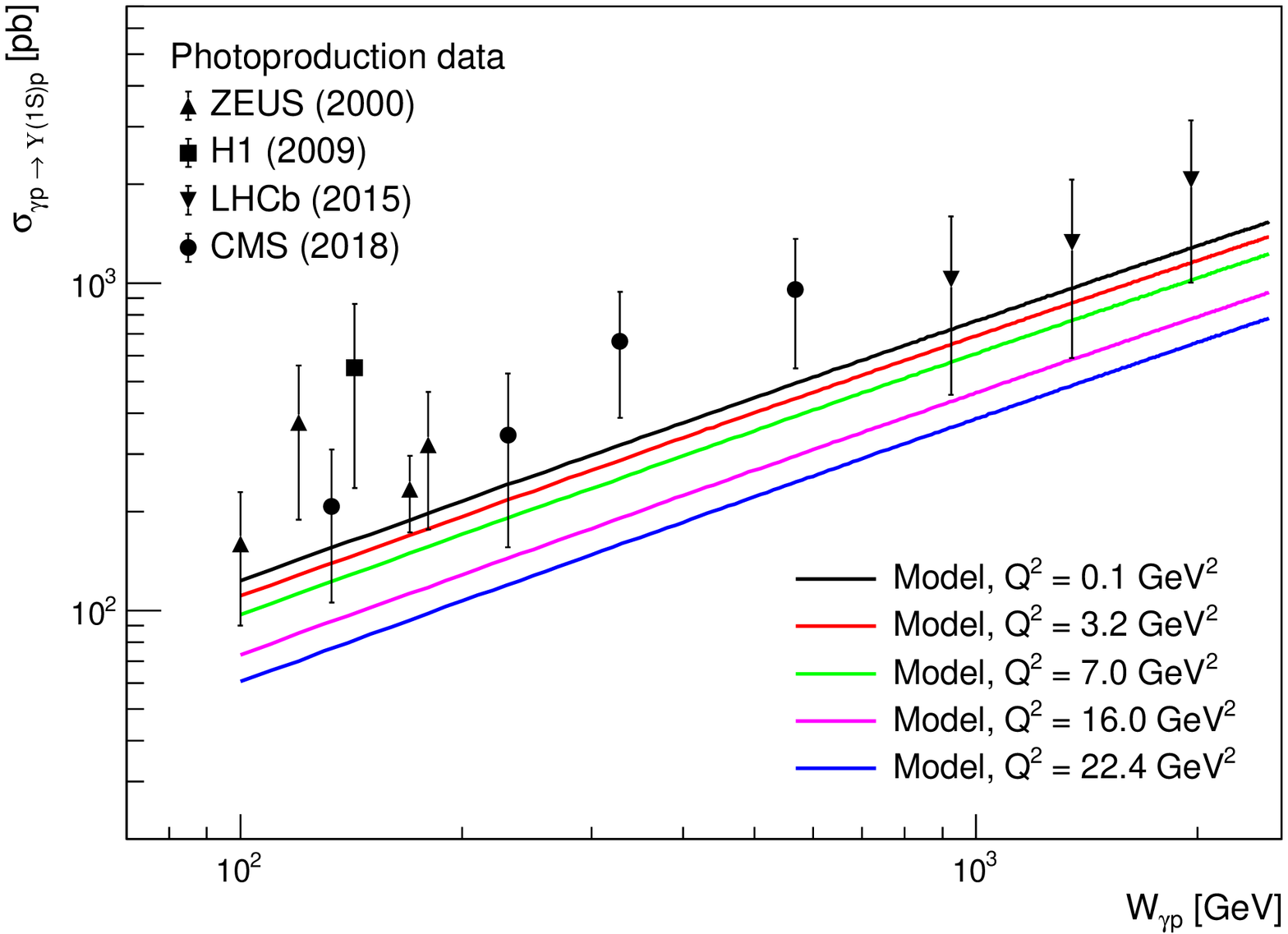}
\includegraphics[width=0.48\textwidth]{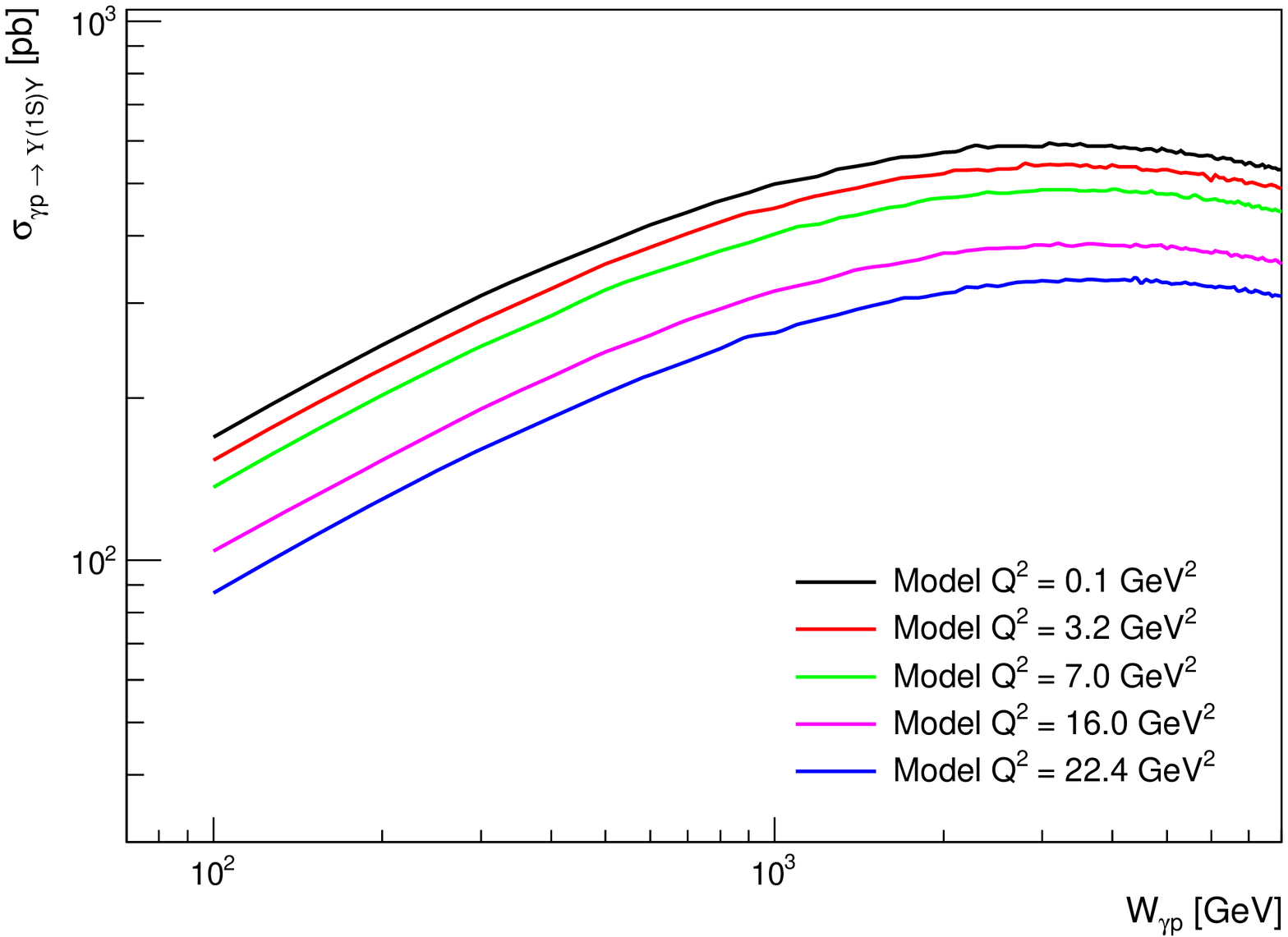}
\caption{\label{fig:upsilon} (Color online) Comparison of the model predictions (solid lines) with H1~\cite{Adloff:2000vm}, ZEUS~\cite{Chekanov:2009zz}, LHCb~\cite{Aaij:2015kea} and CMS data~\cite{Sirunyan:2018sav} for the  $\Wgp$ dependence of the exclusive (left) and dissociative (right) photo- and electroproduction  cross section of a $\Uos$ meson.}
\end{figure*}

The photoproduction of $\jpsi$ has already been studied in~\cite{Cepila:2016uku,Cepila:2018zky}. Here we show the same comparison of H1~\cite{Alexa:2013xxa} and ALICE p--Pb data~\cite{TheALICE:2014dwa} with the model predictions in Fig. \ref{fig:J-psi}. Additionally, recent ALICE data~\cite{Acharya:2018jua} is included. These new photoproduction measurements are also correctly described by the predictions. Electroproduction data from H1~\cite{Aktas:2005xu} is also shown in the figure.  The predictions for the exclusive and dissociative cross sections show a  good agreement with all these data. 

The comparison between the predictions for the exclusive and dissociative photoproduction cross section for the $\Uos$ vector meson and data  has been presented
in~\cite{Cepila:2018zky}. We  present it here again in Fig.~\ref{fig:upsilon} to provide a comparison
with the electroproduction predictions which are the main topic of this work. The exclusive photoproduction cross section is compared with ZEUS~\cite{Adloff:2000vm} and H1~\cite{Chekanov:2009zz} data from HERA. It is also compared with LHCb data taken in proton-proton collisions at  $\sqrt{s} = 7$ TeV  and $\sqrt{s} = 8$ TeV at the LHC~\cite{Aaij:2015kea}. The last set of data we compare our predictions with was measured by
 CMS in p--Pb collisions at  $\sqrt{s} = 5.02$ TeV~\cite{Sirunyan:2018sav}. 
 The data is correctly described by the predictions, although the current uncertainty of the measurement does not allow us to extract strong conclusions regarding the agreement between data and the model.
 Currently, to our knowledge there is no electroproduction data for the exclusive nor for the dissociative process. We expect   these measurements to be performed at  future electron-ion colliders.
 
 \begin{figure*}
\includegraphics[width=0.48\textwidth]{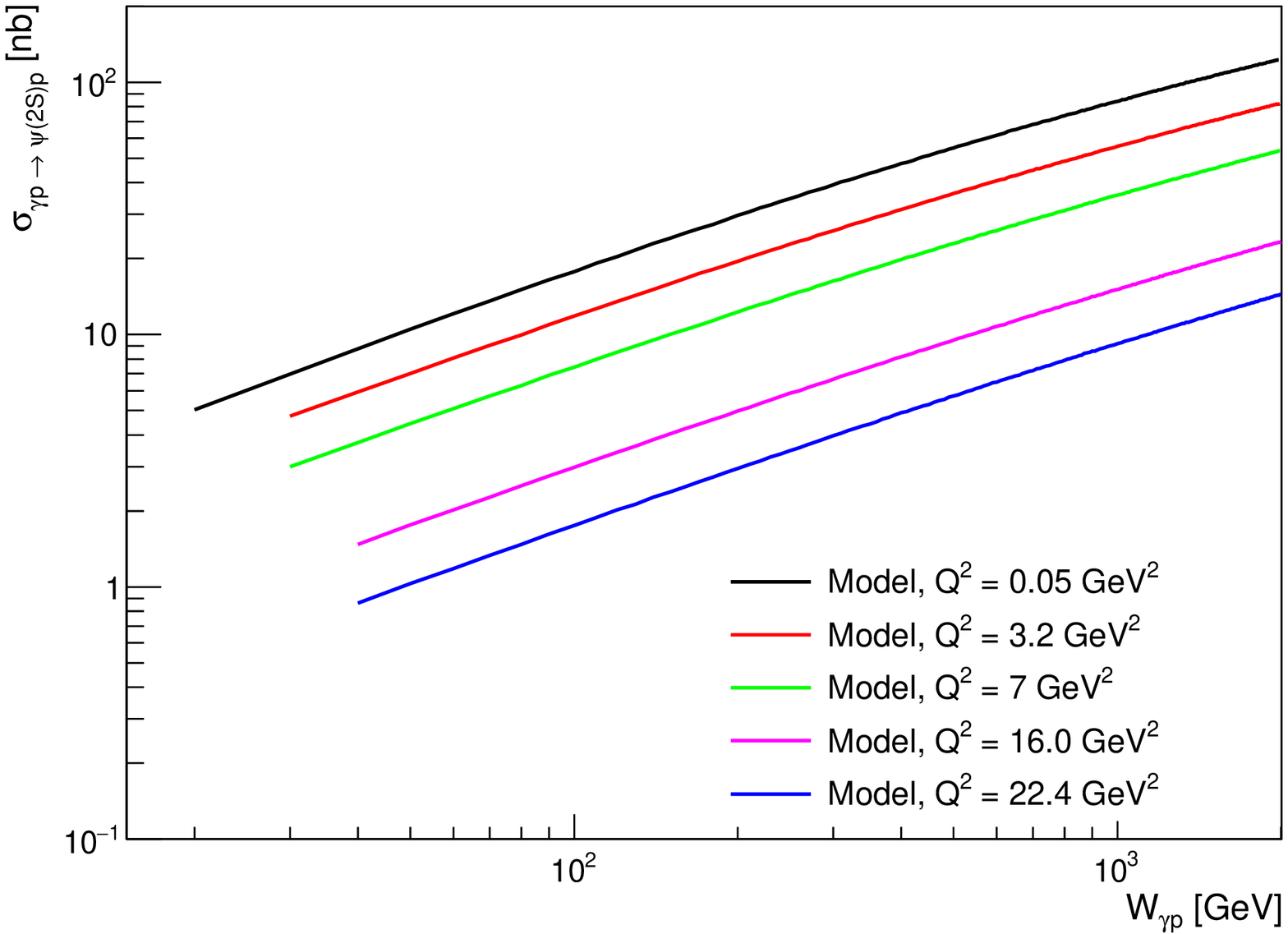}
\includegraphics[width=0.48\textwidth]{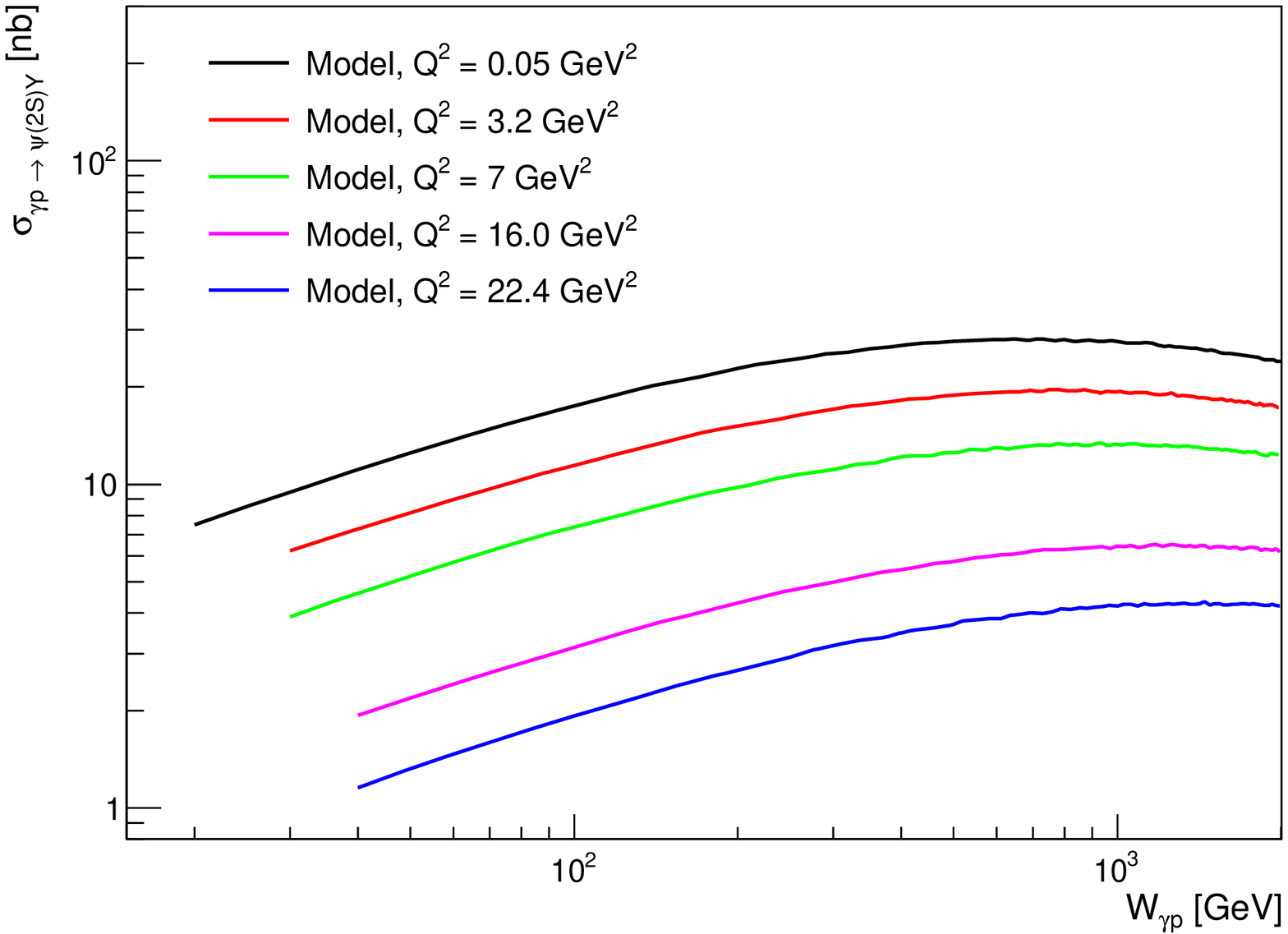}
\caption{\label{fig:psi2S} (Color online) Model predictions (solid lines)  for the  $\Wgp$ dependence of the exclusive (left) and dissociative (right) photo- and electroproduction  cross section of a $\psip$ meson. }
\end{figure*}

\begin{figure*}
\includegraphics[width=0.48\textwidth]{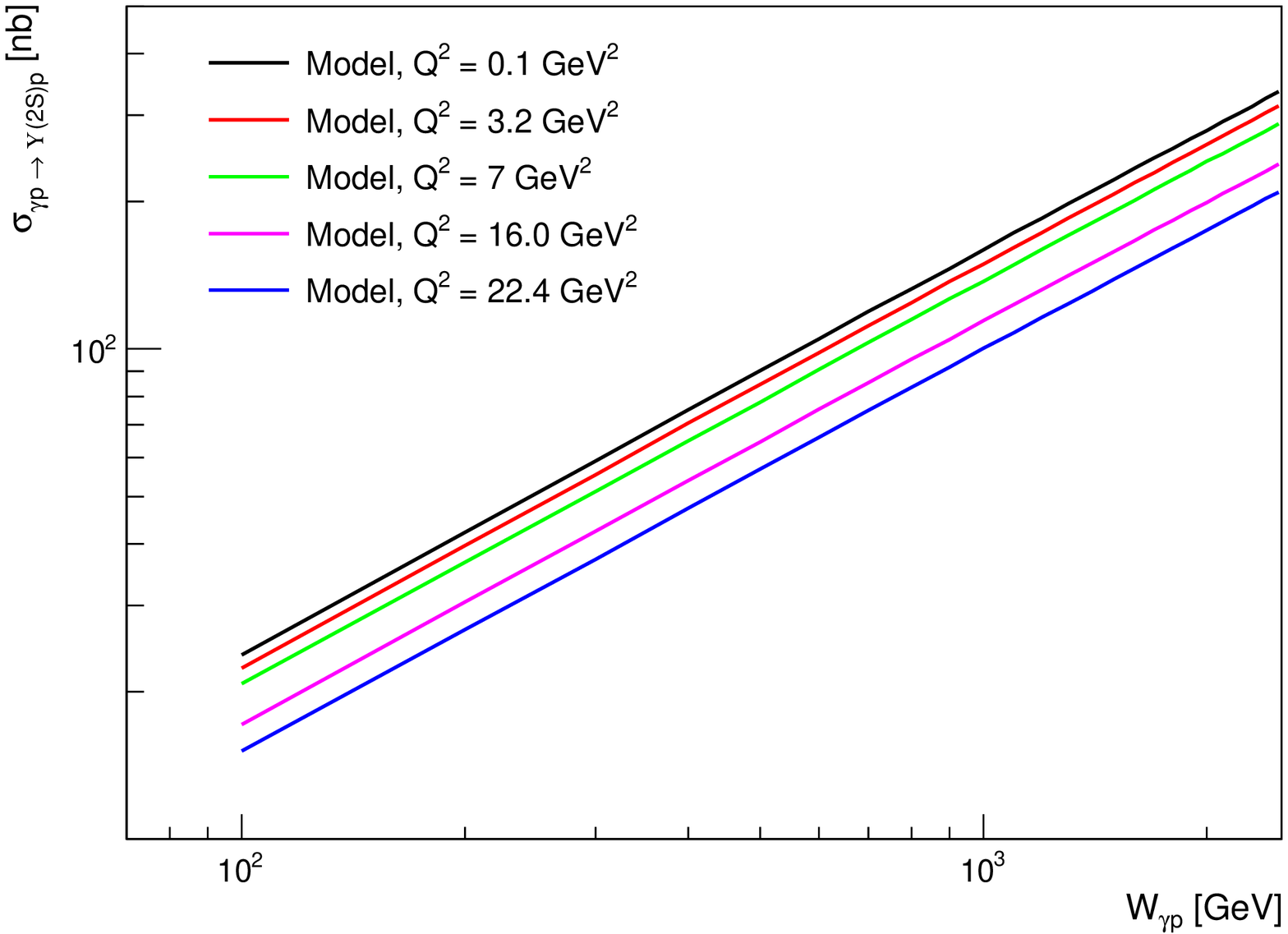}
\includegraphics[width=0.48\textwidth]{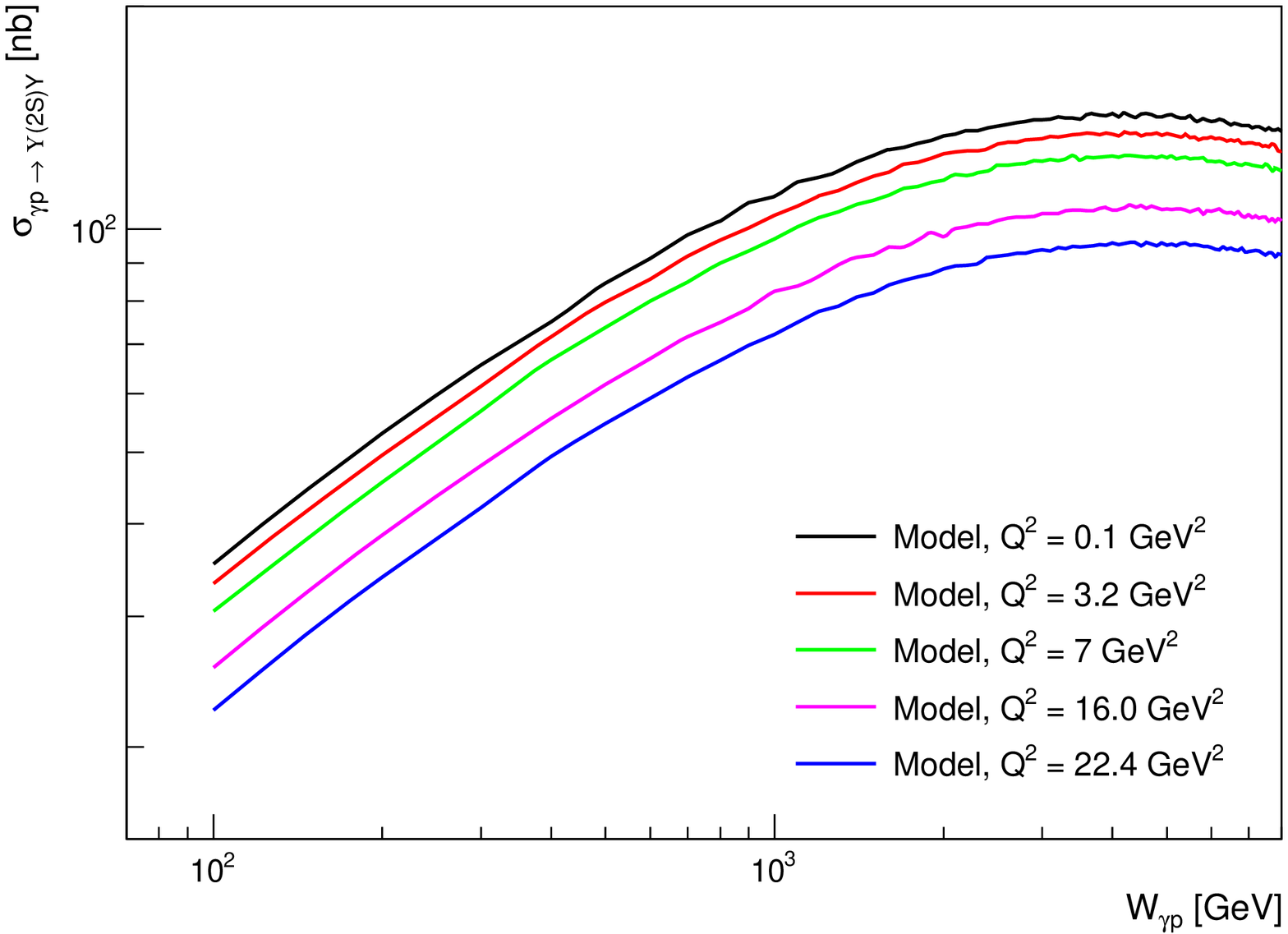}
\caption{\label{fig:upsilon2S} (Color online) Model predictions (solid lines)  for the  $\Wgp$ dependence of the exclusive (left) and dissociative (right) photo- and electroproduction  cross section of a $\Uts$ meson. }
\end{figure*}

To complete the set of our predictions we  present predictions for the excited states $\psip$ and $\Uts$ in Figs.~\ref{fig:psi2S} and~\ref{fig:upsilon2S}. Currently,  there is no direct data for these particles, but there is photoproduction data from H1~\cite{Adloff:2002re} and electroproduction data at $Q^2=16$ GeV$^2$ from ZEUS~\cite{Abramowicz:2016xls} for the ratio of the exclusive production of $\psip$ to that of $\jpsi$. Our predictions describe correctly the measured ratios, 
although, as in the case of the $\Uos$, the current uncertainty of the measurement does not allow us to extract strong conclusions regarding the agreement between data and the model.

In summary, there is a good agreement between all existing data for the exclusive and the dissociative photo- and electoproduction of vector mesons and the predictions of our model.
 
\section{Geometric saturation scale\label{sec:gss}}
\subsection{Introduction of the geometric saturation scale}

As already noticed in~\cite{Cepila:2016uku} for the case of $\jpsi$ photoproduction and confirmed in~\cite{Cepila:2018zky}  for the photoproduction of $\Rz$ and $\Uos$, the behavior of the dissociative cross section as a function of the photon-proton center-of-mass energy is quite striking. At low energies, the cross section rises with $\Wgp$ to reach a maximum, after which it decreases steeply.
The same behavior is observed for the dissociative electroproduction of vector mesons. Interestingly, the position of the maximum depends not only on the mass of the vector meson, but also on the virtuality of the exchanged photon. 

The interpretation of this behavior is given by the form of the cross section shown in Eq.~(\ref{VM-cs-diff-disoc}). The dissociative production measures the variance over the different configurations into which the structure of the proton can fluctuate. In our model, this is given by the different geometrical placements of the hot spots in the impact-parameter plane. As the energy $\Wgp$ increases, so it does the number of hot spots  inside the proton as shown in Eq.~(\ref{eq:Nhsx}). As the hot spots have all the same transversal area, the more hot spots there are, the more the proton area is filled. At some point, all the possible configurations start to look alike, because all of them start filling all the available area in the proton and overlap in a process reminiscent of percolation~\cite{Armesto:1996kt}. From this energy onwards  the variance over configurations steeply decreases.  The maximum of the dissociative cross section defines a well defined energy at a well defined scale. We call this point the {\em geometric saturation scale} (GSS) and in the following study some of its properties. 
 
\subsection{Energy dependence of the geometric saturation scale}

\begin{figure*}
\includegraphics[width=0.48\textwidth]{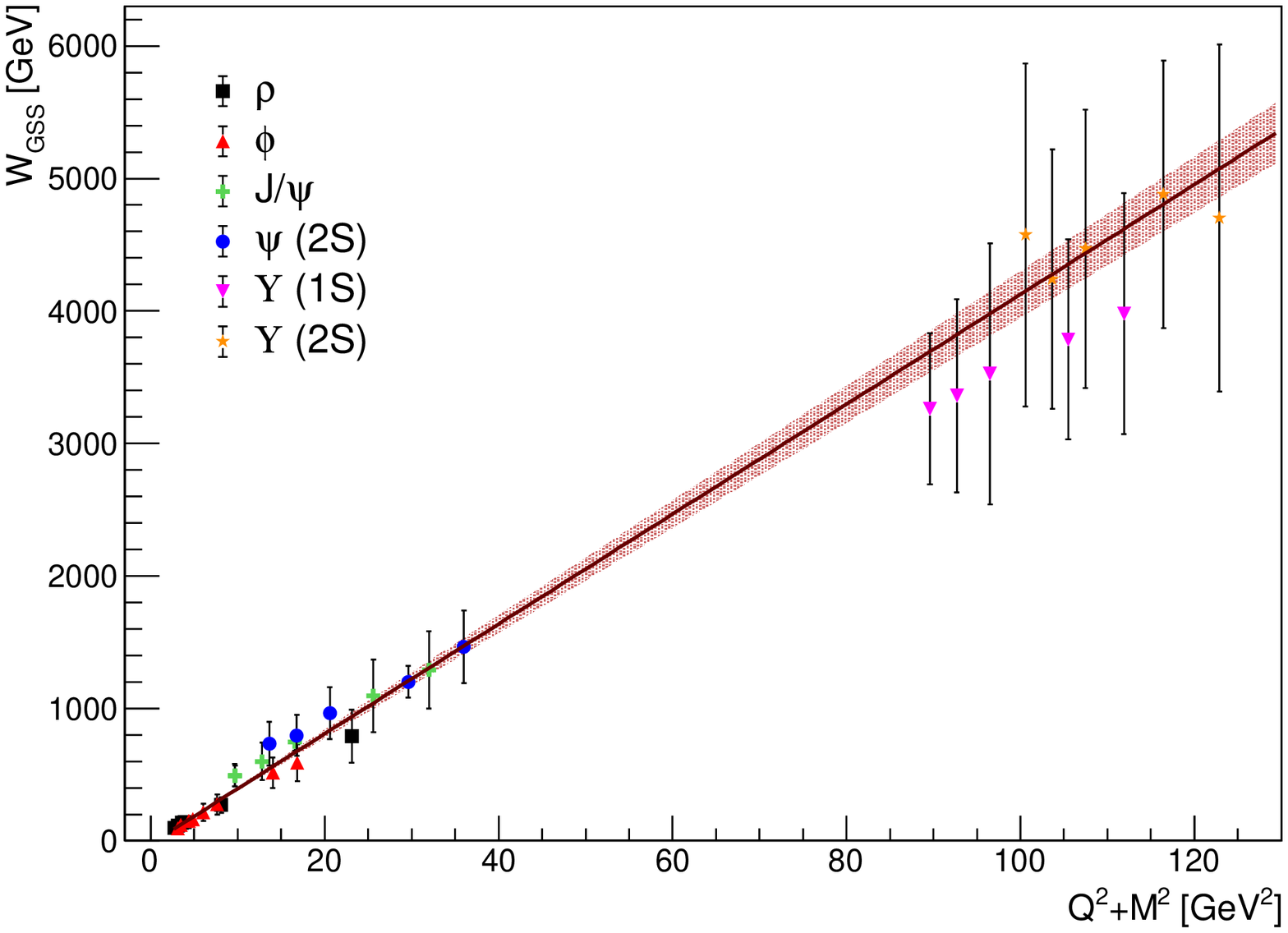}
\includegraphics[width=0.48\textwidth]{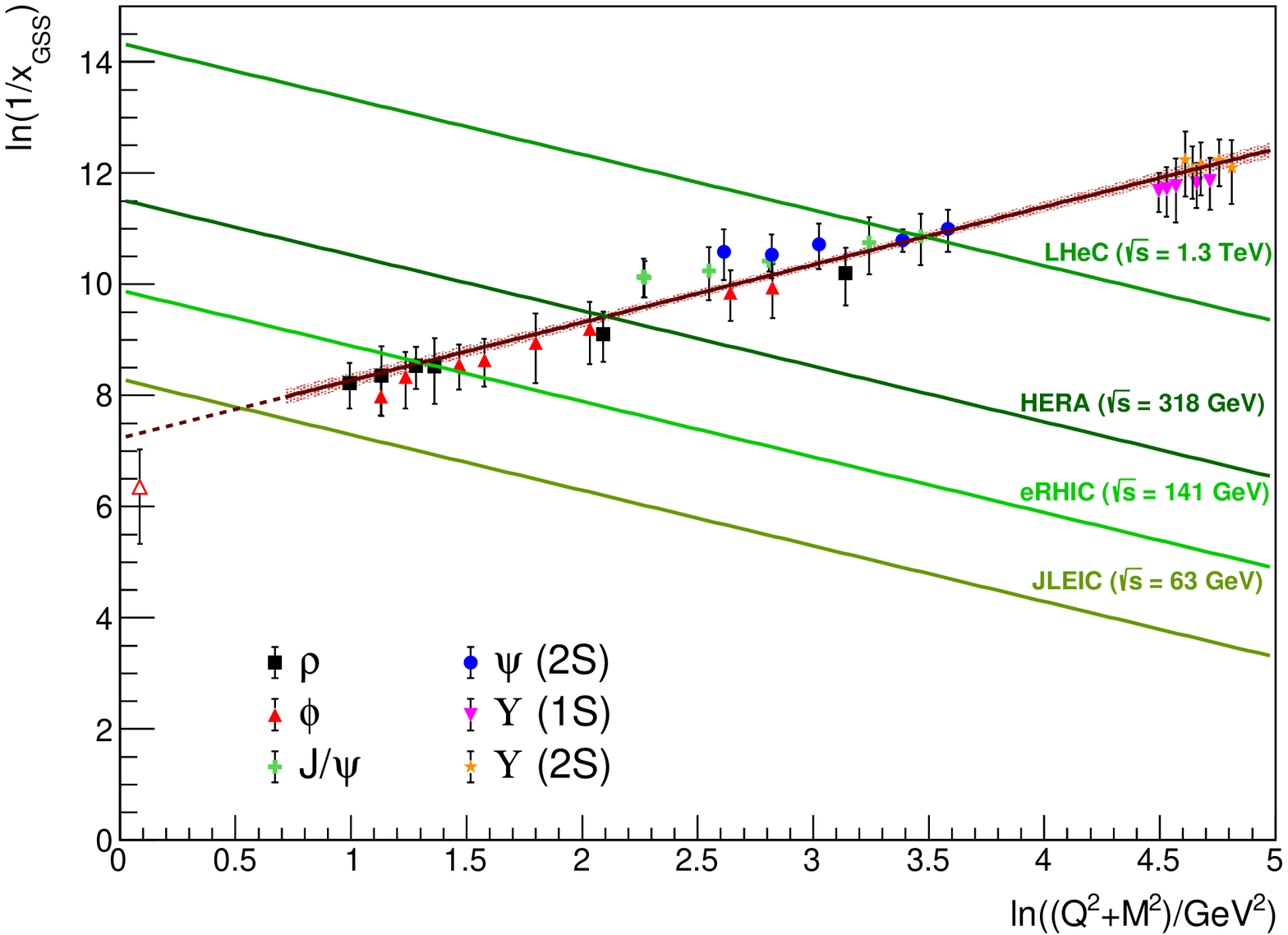}
\caption{\label{fig:mass-Q_dep} (Color online) Left: Position of the maxima of the dissociative cross sections (markers) and an estimation of the related uncertainty (bars) as a function of $Q^2+M^2$. The line is a fit to the line defined in Eq.~(\ref{eq:Wgss}) and the band represents the one sigma contour.
Right: the same data as in the left panel, but translating $W_{\rm GSS}$ into $x_{\rm GSS}$ and  plotting them in logarithmic variables. The red line is the fit to Eq.~(\ref{eq:xgss}) The diagonal lines represent the kinematic reach of some of the proposed future electron-ion colliders. See text for details.}
\end{figure*}

For each of the vector mesons and for each of the virtualities we determine the energy $W_{\rm GSS}$ at which the maximum is found. As the predictions are based on a random process, the value at the maximum may fluctuate a bit, so we chose a region containing the 1\% largest values of the cross section to determine the position of the maximum along with an estimation of the associated uncertainty.

Figure~\ref{fig:mass-Q_dep} shows in the left panel the position of the maximum as a function of $Q^2+M^2$, which is a  measure of the scale of the process. The behavior seems to be  linear, so we fitted the extracted maxima to the functional form
\begin{equation}
W_{\rm GSS} = a_0 + a_1(Q^2+M^2).
\label{eq:Wgss}
\end{equation}
For the fit we considered only points with $Q^2+M^2$ larger than 2 GeV$^2$.  The fit is good. The $\chi^2$ per degree-of-freedom is 0.41, the small value reflecting the large assigned uncertainty on the position of the maxima. The parameter values are $a_0=-21\pm11$ GeV and the slope that is obtained is $a_1=41.5\pm1.8$ GeV$^{-1}$.

Using Eq.~(\ref{x}) we can translate  $W_{\rm GSS}$ into $x_{\rm GSS}$. The result is shown in the right panel of Fig.~\ref{fig:mass-Q_dep}. The behavior is also linear in the logarithmic variables so we have fit the predictions to 
\begin{equation}
\ln(1/x_{\rm GSS}) = c_0 + c_1\ln((Q^2+M^2)/{\rm GeV}^2).
\label{eq:xgss}
\end{equation}
We found $c_0 = 7.2\pm0.2$ and $c_1=1.04\pm0.06$. The same figure shows the kinematic limit of some of the proposed future electron-ion colliders. This limit is obtained from 
\begin{equation}
xys=Q^2,
\end{equation}
where the inelasticity of the collision is set to $y=1$ and the center-of-mass energies $s$ of the accelerators are obtained from the energies of the proton, $E_p$, and electron, $E_e$,  beams taken from  Tab. I of~\cite{Lomnitz:2018juf}: $E_e =10$ GeV, $E_p= 100$ GeV for JLEIC; $E_e =18$ GeV, $E_p= 275$ GeV for eRHIC; $E_e =27.5$ GeV, $E_p= 920$ GeV for HERA; and $E_e =60$ GeV, $E_p= 7$ TeV for LHeC.

It is interesting to notice that even for the collider with the lower energy, one could measure this linear behavior using electroproduction of $\Rz$ and of $\phi$ vector mesons at relatively small virtualities, but in all cases at scales $Q^2+M^2$ above 1 GeV$^2$. The  detectors at the JLEIC and eRHIC are  still under development, but the envisaged capabilities would allow the measurement of $\Rz$ and $\phi$ as discussed in detail in~\cite{Lomnitz:2018juf}.  To investigate the positions of the maxima for $\jpsi$ one needs the LHC and the LHeC for photo- and electroproduction cases, respectively. The positions of the maxima for the Upsilon states seems to be out of reach even for the LHeC.

\section{Summary and outlook\label{sec:sum}}

Using the energy-dependent hot spot model we have presented predictions for the exclusive and dissociative electroproduction of vector mesons off proton targets. We studied the production of $\Rz$, $\phi$, $\jpsi$, $\psip$, $\Uos$ and $\Uts$ states. We found that the dissociative cross section as a function of $\Wgp$ presents a maximum and have used this maximum to define a geometrical saturation scale. We found that the energy evolution of this scale is linear in $Q^2+M^2$ and that this behavior can be studied at the planned JLEIC, eRHIC and LHeC electron-ion colliders.

To be able to perform such measurements the detectors would have to be instrumented in the forward rapidity regions in order to tag the presence of the products from the dissociative state. Such a technique has been used at HERA in the past; it is also used nowadays at the LHC to reject the dissociative events when measuring the exclusive production channel, so it seems to be feasible if planned in advance.

Mapping the energy evolution of the geometric saturation scale provides an extra handle to investigate quantitatively the high-energy limit of QCD and to study the phenomenon of gluon saturation in the proton.

\section*{Acknowledgements}
We thank Victor Gon\c{c}alves for useful discussions.
This work has been partially supported by the following grants:
17-04505S of the Czech Science Foundation (GA\v{C}R),  COST Action CA15213 THOR,
 LTC17038 of the INTER-EXCELLENCE program at the Ministry of Education, Youth and Sports of the Czech Republic and the European Regional Development Fund-Project 
"Brookhaven National Laboratory - participation of the Czech Republic" 
(No. CZ.02.1.01/0.0/0.0/16\_013/0001569).

\bibliography{ALICE,QCD,HERA,CMS,LHCothers}

\end{document}